# Bus Trajectory-Based Street-Centric Routing for Message Delivery in Urban Vehicular Ad hoc Networks

Gang Sun[1,2], Yijing Zhang[1], Dan Liao[1], Hongfang Yu[1,2], Xiaojiang Du[3], Mohsen Guizani[4]

[1]Key Lab of Optical Fiber Sensing and Communications (Ministry of Education), UESTC, Chengdu, China
[2]Center for Cyber Security, UESTC, Chengdu, China
[3]Department of Computer and Information Sciences, Temple University, Philadelphia, PA, USA
[4]Department of Electrical and Computer Engineering, University of Idaho, Moscow, Idaho, USA

*Abstract*—This paper focuses on the routing algorithm for the communications between vehicles and places in urban VANET. As one of the basic transportation facilities in an urban setting, buses periodically run along their fixed routes and widely cover city streets. The trajectory of bus lines can be seen as a sub map of a city. Based on the characters of bus networks, we propose a bus trajectory-based street-centric routing algorithm (BTSC), which uses bus as main relay to deliver message. In BTSC, we build a routing graph based on the trajectories of bus lines by analyzing the probability of bus appearing on every street. We propose two novel concepts, i.e. the probability of street consistency (PSC) and the probability of path consistency (PPC) which is used as metrics to determine routing paths for message delivery. This aims to choose the best path with higher density of busses and lower probability of transmission direction deviating from the routing path. In order to improve the bus forwarding opportunity, we design a bus-based forwarding strategy with ant colony optimization (FACO) to find a reliable and steady multi-hop link between two relay buses in order to decrease end-to-end delay. BTSC makes the improvements in the selection of routing path and the strategy of message forwarding. Simulation results show that our proposed routing algorithm has a better performance in transmission ratio, transmission delay and adaptability to different networks.

## I. INTRODUCTION

AS the development of the intelligent transportation system (ITS), vehicular ad hoc network (VANET) is excepted to support more and more new applications to improve the qualities of people's life and traffic, such as location-based services, travel planning services, traffic accident reporting services, intelligent parking services and social vehicular platforms [1-4]. Acting as mobile intelligent communication equipment, vehicles are expected to access the network wherever and whenever possible. There have been several papers focused on security [5-8] and mobility topics [9,10], but efficiency also plays an important role in VANET. Thus it's significant to design efficient routing algorithm in VANET. In this paper, we will focus on the design of routing algorithms for the communication between vehicles and some places.

In VANET, vehicles equipped with on-board units (OBUs) can spontaneously form a self-organization network without depending on other infrastructures [11-14]. Vehicles can directly communicate with other vehicles or infrastructures within their communication range by using dedicated short range communication (DSRC) technology or indirectly communicate with others by multi-hop links [11,15-18]. Intermediate vehicles between sender and destination play the role of router and deliver packets by using the carry-and-forward mechanism [17,19-23]. On one hand, the movement of vehicles is affected by the driver's subjective awareness, causing the uneven distribution and unpredictable trajectories of vehicles [5,21]. Dense network is beneficial for multi-hop link and has a good routing performance. However, in sparse network, it's difficult to find the relay to deliver packets. Consequently, routing performance becomes poor [23-24]. On the other hand, the movement of vehicles makes rapid and frequent change of the network's topology [25-26]. Therefore, the conventional routing based on the technology of ad hoc network has poor performance [26-34].

As one of the infrastructures in VANET, road side units (RSUs) can work as the router to store and forward packets [26,28,35]. However, the performance of RSU-based routing is limited by the number and also by the location of RSUs [23]. Only the intensive deployment of RSU can maximize the availability of RSU and provide a stable service for message transmission in VANET. Otherwise the link between vehicles may disconnect in the blind area of communication [36]. However, the deployment and management of RSU cause a lot of overhead, and the limited communication range and fixed position cause short connection between RSU and moving vehicles [23,36-37]. Therefore, the RSU-based routing is inapplicable and uneconomical.

As one of the basic transportation facilities, buses pervade the main roads in city, and have specific driving trajectories and departure intervals [11,37]. Compared with the common vehicular network, the bus network has the characteristics of wide coverage, relatively uniform distribution of nodes, fixed trajectory and regular service [38-39]. Due to the unique advantage of buses, many researchers have paid attention to the superiority of bus network, and proposed some bus-based routing algorithms to solve the aforementioned routing problem [15,23,38,40,43]. Literature [23] and [40] built the bus-based backbone graph for routing. Literature [41] used taxis and buses as the communication backbone. Literatures [15] and [42] proposed the bus-based two-tier VANET architecture, where buses are the only relay to forward packets. On one hand, the existing bus-based routing algorithms are all node-centric: the routing path is a multi-hop link which consists of some buses between sender and destination, and once the routing plan is made, only specified buses are used to forward packets in sequence. The carrier of packet can forward packets to the next only when it encounters with the specified next relay bus. And buses in different bus lines can encounter only in the overlap of the trajectories of their bus lines. When carriers miss the specified next relay bus in their meeting place, it will take a long time before the next encounter, which happens quite often in practice. On the other hand, in those existing bus-based routing strategy, bus is the only relay to deliver packets and ordinary vehicle only serve as a sender or receiver. Therefore, bus undertakes all the routing tasks in the network, and the number of buses is less than the ordinary vehicle's, which may cause network congestion.

To solve the above problems, in this paper, we proposed the bus-based street-centric routing strategy (BTSC) with considering that the movement of vehicles is restricted by the topology of streets. Firstly, we build the bus lines-based routing graph used to select routing trajectory by analyzing the relationship between the trajectory of bus lines and streets. Secondly, based on our proposed routing graph, we select a sequence of streets with high density of bus as the routing path. Buses along the routing path work as the relay to delivery packets to the destination by using the mechanism of carry-and-forward. In the carry-and-forward mechanism, the amount of time to carry information is far greater than the amount of time to forward information, therefore, the time of carrying information is the main factor of influencing the performance of routing. Then, we propose the bus-based forwarding strategy with ant colony optimization (FACO). This strategy implements ACO algorithm to find the optimal next relay bus when the carrier of packets cannot find next available relay in its communication range, aiming at decreasing the time of carrying during routing. The main contributions of this paper are listed as follows.

1) We propose the bus line-based routing graph. By analyzing the trajectory of bus lines, every edge in routing graph is assigned a weight to reflect the density of bus on the street corresponding to the edge.
2) We propose two novel concepts called the probability of street consistency (PSC) and the probability of path consistency (PPC). The former is used to describe the consistency of bus lines between two adjacent streets, and the latter is used as the metric to select the routing path with high density of buses.
3) We design the bus-based forwarding strategy with ant colony optimization (FACO). During the process of routing, buses serve as the main relay to deliver packets from source to destination, and ordinary vehicles works as the secondary relay to build a multi-hop link between two buses. Ant colony optimization algorithm is used to find an optimal next relay bus and a stable multi-hop link between two relay buses, aiming at increasing the chance of forwarding and decreasing the delay of routing.
4) We carry out extensive simulations and analyze the average end-to-end delay packet transmission ratio of our routing algorithm. The results of simulations show that our proposed routing algorithm has good performance in the delay and packet loss rate.

The remainder of this paper is as follows. Section 2 introduces the related work. Section 3 presents the preliminaries, including the system model and the definition of link life time. Section 4 shows the detail of BTSC, which consists of three parts: the bus lines-based routing graph, the selection of routing path and the bus-based forwarding strategy with ACO (FACO). Section 5 describes the simulation environment and the analyses of simulation results. Section 6 concludes this paper.

## II. RELATED WORK

### A. Bus-based Routing Protocol

In the existing architecture of Bus-based VANET, bus is not only a communication node, but also a router. Some exiting works build the contact relationship between bus lines by analyzing the information of bus routes or digging the historical track information of bus. Based on the contact relationship, a sequence of bus lines is selected to relay information from source node to destination node. Buses on the selected bus lines will deliver information in a way of store-carry-forward. In contrast, the amount of time to carry information is far greater than the amount of time to forward information during routing. Therefore, the key to improving routing performance in bus-based routing is to increase the frequency of contact between relay buses.

In [15], Jiang et al. proposed a two-tier bus-based VANETs architecture named BUSVANET, in which common cars (namely non-bus) form the low tier and the high tier consist of bus and RSU. All the high-tier nodes (including moving bus and fixed RSU) dynamically form interconnected topology. Packets are delivered between high-tier nodes until the high-tier node with packets near the destination vehicle, and then are forwarded to destination by the high-tier node, which realizes communication between vehicles.

In [23], Zhang et al. proposed a bus-based routing scheme in the VANETs without RSU, in which bus system works as routing backbone. By collecting and analyzing the actual trajectories of buses in Beijing, authors modeled the contact graph of bus lines which reflects the overlapping relationship of bus lines, then established a community-based bus system by using community detection techniques to divide bus lines into different community over the contact graph. In the community-based bus system, bus lines in the same community will contact each other frequently. Based on the routing backbone, the process of routing is divided into the inter-community routing and the intra-community routing. A sequence of bus lines in which buses frequently contact each other is selected preferentially as the relay, thus reducing delay of routing and increasing the delivery ratio.

In [38], Zhang *et al.* studied on the geocast in bus-based VANETs and proposed a geocast routing scheme named Vela. By mining the historical trajectory of bus, authors analyzed the regularity of bus travel-time in temporal and the mode of bus encounter in spatial and built spatial-temporal patterns. Based on these spatial-temporal patterns of bus, Vela provides feasible routing paths with the best possible QoS for data delivery requests.

In [36], Chang *et al.* modeled the bus contact graph by analyzing the trajectories of bus lines and extracted a routing graph from the contact graph according to the position of packet recipient. For every packet, a routing graph is made instead of a routing path to adapt to dynamic urban traffic.

However, those bus-based routing algorithms in these exiting works have two main demerits: 1) excessively depending on bus, only buses have the ability to relay, which may cause network congestion because the number of buses is much smaller than the number of common cars; 2) While adopting the mechanism of node-centric, the routing path would consist of specified nodes, which decreases the chance to forward packets and cause longer delay.

### B. Street-Centric Routing Protocol

In VANETs, routing protocols have two categories: the node-centric routing protocol and the street-centric routing protocol. In the former mechanism, the routing path consists of some moving vehicles. But the latter mechanism uses a sequence of streets as the routing trajectory. In practice, the trajectories of vehicles are some sequence of streets, and the movement of vehicles is restricted by the road topology and traffic conditions. Thus, the topology of the network can be seen as a subset of city maps. Some exiting works have proved that the street-centric routing mechanism has better performances than the node-centric routing mechanism [44]. The street-centric routing protocol mainly consists of two aspects: the selection of routing path and the forwarding strategy. The routing path is a sequence of streets, and the forwarding strategy is designed for relay vehicles to deliver packets from its sender to its recipient along the routing path.

In [45], authors proposed a new concept named Micro topology (MT), which is made up of moving vehicles along streets and wireless links between vehicles. Considering the end-to-end routing performance in a MT, authors designed a street-centric routing protocol based on MT, which includes selecting the next MT for routing and designing packets delivery strategy within a MT. Within a MT, packets are ferried from the start to the end by intermediate vehicles. When packets arrive at the end of the MT, protocol will dynamically select a new MT as the next routing MT with consideration of avoiding routing loops in routing decisions.

In [46], Zhang *et al.* utilized a Wiener process to predict link probability between vehicles and built a link model. Based on this model, routing streets are selected dynamically according to the link status. To decrease cost and increase the reliability of routing, authors proposed a novel concept called ETCoP as the metric to choosing relay nodes.

In [47], Kadadha *et al.* proposed a novel optimized link state routing protocol for urban VANET, which is based on QoS to select multipoint relays. And the street-centric parameters (such as lane weight) were used as the metric of choosing relays for the first time.

Note that the trajectory of bus lines is also street-centric, and bus lines have a relationship with streets, which can be utilized in the design of routing protocol for achieve better routing performance. However, those existing works about the street-centric routing all fail to utilize the relationship between streets and bus lines.

### III. PRELIMINARIES

#### A. System Model

In this paper, we propose a two-layer architecture of VANET (as shown in Fig. 1), where vehicle nodes (which consist of buses and common cars) are divided into two layers, i.e. upper layer and low layer. Both buses and common cars (namely non-bus) can send, receive and relay packets in our proposed architecture of VANET, which is different from the existing bus-based VANET architectures where only buses can relay packets.

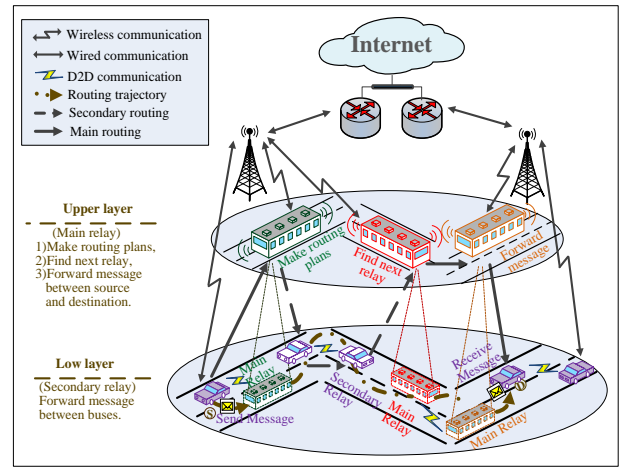

Fig.1. System architecture

● Upper layer: consists of all the bus nodes which are equipped with on-board units (OBUs), GPS, digital map and the information of bus lines in city and on-board processing units which is larger than common car's. As the main relay, buses can make the plan of routing path for packets by finding the next available relay bus for packets and forward packets.

● Low layer: consists of all the common cars which are equipped with on-board units and GPS. As the secondary relay, common cars are used to forward packets between two main relays which cannot directly communicate with each other because of distance. When the main relay cannot find the next available main relay in its communication range, it can find the next eligible and available relay bus and multi-hop links between them with the help of secondary relays.

Every vehicle has the ability to communicate with other vehicles which are within its communication range by using dedicated short-range communication technology (DSRC) and periodically broadcasts beacon messages (including ID, position, velocity and moving direction). In addition, every vehicle has a table (called NeighborTable) to save the information of its neighbor vehicles when it receives the beacon message from the neighbor vehicle. To maintain NeighborTable, every vehicle will periodically delete overdue information of vehicles from NeighborTable.

*B. Link Life Time*

We assume that the communication radius of vehicles is R. For any two vehicles $C_i$ and $C_j$, the position and velocity of $C_i$ and $C_j$ are $\boldsymbol{P_1}, \boldsymbol{V_1}, \boldsymbol{P_2}$ and $\boldsymbol{V_2}$ respectively, and the velocity difference and the distance between $C_i$ and $C_j$ are respectively denoted by $\Delta D$ and $\Delta V$. Vectors $\boldsymbol{P_1}, \boldsymbol{V_1}, \boldsymbol{P_2}, \boldsymbol{V_2}, \Delta D$ and $\Delta V$ are represented as follows.

$$\mathbf{P_1} = \begin{bmatrix} x_1 \\ y_1 \end{bmatrix}, \mathbf{V_1} = \begin{bmatrix} v_{x_1} \\ v_{y_1} \end{bmatrix}, \mathbf{P_2} = \begin{bmatrix} x_2 \\ y_2 \end{bmatrix}, \mathbf{V_2} = \begin{bmatrix} v_{x_2} \\ v_{y_2} \end{bmatrix},$$

$$\Delta \mathbf{D} = \mathbf{P_2} - \mathbf{P_1} = \begin{bmatrix} x_2 - x_1 \\ y_2 - y_1 \end{bmatrix}, \Delta \mathbf{V} = \mathbf{V_2} - \mathbf{V_1} = \begin{bmatrix} v_{x_2} - v_{x_1} \\ v_{y_2} - v_{y_1} \end{bmatrix}.$$

The distance between $C_i$ and $C_j$ changes over time, which is given as follows.

$$\begin{aligned} D(t)^2 &= [x_2(t) - x_1(t)]^2 + [y_2(t) - y_1(t)]^2 \\ &= |\Delta \mathbf{V}|^2 \times t^2 + 2 \cdot (\Delta \mathbf{D} \square \Delta \mathbf{V}) \times t + |\Delta \mathbf{D}|^2 \end{aligned} \quad (1)$$

If $D(t)^2 \leq R$, $C_i$ is within the communication range of $C_j$ it can then communicate with $C_j$ directly at time $t$. If the initial distance $|\Delta D| \leq R$, the duration of connection between $C_i$ and $C_j$ is denoted by $T(l_{i,j})$ and its calculation formula is as follows.

$$T(l_{i,j}) = \begin{cases} 0, & \text{if } |\Delta \mathbf{D}| = R, \cos(\Delta \mathbf{D}, \Delta \mathbf{V}) \geq 0 \\ \infty, & \text{if } \Delta \mathbf{V} = 0 \\ \dfrac{-B + \sqrt{B^2 - 4 \times A \times C}}{2 \times A}, & \text{otherwise} \end{cases} \quad (2)$$

where,

$$\begin{cases} A = |\Delta \mathbf{V}|^2 \\ B = 2\Delta \mathbf{D} \square \Delta \mathbf{V} \\ C = |\Delta \mathbf{D}|^2 - R^2 \\ \cos(\Delta \mathbf{D}, \Delta \mathbf{V}) = \dfrac{\Delta \mathbf{D} \square \Delta \mathbf{V}}{|\Delta \mathbf{V}| \times |\Delta \mathbf{D}|} \end{cases} \quad (3)$$

If these two vehicles move at the same velocity (namely the equivalent value and the same direction of velocity), they can communicate with each other until their velocity changes, that is denoted by $T(l_{i,j}) = \infty$. If the value of initial distance $|\Delta D|$ equals R and the angle between the initial velocity difference $\Delta V$ and the initial distance $\Delta D$ varies from $-\frac{\pi}{2}$ to $\frac{\pi}{2}$, $C_i$ disconnect with $C_j$ at the beginning, namely $T(l_{i,j}) = 0$.

During the communication between $C_i$ and $C_j$, the velocities of $C_i$ and $C_j$ obey Gaussian distributions which are represented as $v_i \sim N(\mu_i, \sigma_i^2 t)$ and $v_j \sim N(\mu_j, \sigma_j^2 t)$ respectively. And the velocity difference between these two vehicles follows the Gaussian distribution which is represented as $\Delta v \sim N(\mu_i - \mu_j, (\sigma_i^2 + \sigma_j^2)t)$.

According the definition of link reliability in [46], link reliability refers to the probability that the direct communication between two vehicles will continue to be available in a specified period, which is given as follows.

$$r_t(l_{i,j}) = \begin{cases} \int_t^{t+T(l_{i,j})} f(T) dT, & \text{if } T(l_{i,j}) > 0 \\ 0, & \text{otherwise} \end{cases} \quad (4)$$

where,

$$f(T) = \frac{2R}{\sqrt{2\pi}\sigma T^2} e^{-\frac{(\frac{2R}{T} - \mu)^2}{2\sigma^2}}, \quad \text{if } T \geq 0 \quad (5)$$

Where $\mu$ and $\sigma$ respectively represent the expectation and variance of the velocity difference between these two vehicles (namely $\Delta v$), so that $\mu = \mu_i - \mu_j$ and $\sigma = \sqrt{\sigma_i^2 + \sigma_j^2}$.

According the definition in [46], the expected life time of the link between these two vehicles is given as follows.

$$LT(l_{i,j}) = r_t(l_{i,j}) \times T(l_{i,j}) \quad (6)$$

## IV. ALGORITHM DESIGN

In this section, we introduce our proposed bus trajectory-based street-centric routing algorithm (BTSC) which is one of the street-centric routing approaches. The principal features of street-centric routing algorithm consist of two aspects: 1) making a plan of routing path that is a sequence of streets actually; 2) routing packet along the routing path with the help of vehicles. In the process of routing, vehicles use the carry-and-forward mechanism to delivery packets. In this paper, buses are the main relays to delivery packets from sender to destination, and common cars work as the secondary relays used to build multiple-hop link between two main relay which cannot communicate with each other directly. Considering routing failure caused by the sparse network along the routing path, we can estimate the density of bus on the routing path by analyzing the trajectory of bus lines because buses have fixed driving trajectories. Our proposed routing algorithm is mainly made up of three parts: building a routing graph based on the trajectory of bus lines, selecting a routing path and forwarding packets along the routing path.

*A. Bus Lines-Based Routing graph*

In an urban setting, buses pervade the main roads in city and the trajectory of all bus lines can be seen as a sub map. By analyzing the trajectory of bus lines, we discover the correlation between bus lines and streets, and define the probability of a bus appearing on the street as shown below.

*Definition* 1: If the trajectory of bus $b$ includes street $r$, the probability of bus $b$ appearing on street $r$ is the ratio of the length of street $r$ to the total length of the trajectory of bus $b$, namely

$$P_b(r) = \frac{L_r}{L_b} \times f_b(r) \quad (7)$$

where

$$f_b(r) = \begin{cases} 1, & \text{if bus } b \text{ passes street } r \\ 0, & \text{otherwise} \end{cases} \quad (8)$$

Symbols $L_r$ and $L_b$ respectively represent the length of street $r$ and the length of the trajectory of bus $b$. All the streets make up a set called **S**. $P_b(r)$ denotes the probability of bus $b$ appearing on street $r$, which meets three criteria listed as follows.

$$\begin{cases} 0 \leq P_b(r) \leq 1 \\ \sum_{r \in S} P_b(r) = \dfrac{\sum_{r \in S}(L_r \times f_b(r))}{L_b} = 1 \\ P_b(r_i \cup r_k) = P_b(r_i) + P_b(r_k) \end{cases} \quad (9)$$

*Definition* 2: the probability of buses appearing on street $r$ is derived as follows.

$$P_r = \frac{1}{N_{BUS}} \sum_{b \in \mathbf{B}_r} P_b(r) = \frac{1}{N_{BUS}} \sum_{b \in \mathbf{B}_r} \frac{L_r}{L_b} f_b(r) \quad (10)$$

Where $N_{Bus}$ and $\mathbf{B}_r$ respectively represent the total amount of bus lines and the collection made up by all the bus lines which pass through street $r$. $P_r$ denotes the probability of buses appearing on street $r$, and it reaches three conditions shown as follows.

$$\begin{cases} 0 \leq P_r \leq 1 \\ \sum_{r \in S} P_r = \dfrac{1}{N_{BUS}} \times \sum_{r \in S} \sum_{b \in \mathbf{B}_r} P_b(r) = 1 \\ P_{r_i \cup r_k} = P_{r_i} + P_{r_k} \end{cases} \quad (11)$$

Converted from the real road map, $\mathbb{G} = (\mathbb{V}, \mathbb{E}, \mathbb{W})$ represents the bus lines-based routing graph proposed in this paper, where $\mathbb{V}$, $\mathbb{E}$ and $\mathbb{W}$ respectively represent the set of vertices, the set of edges and the set of weights. According to the probability of buses appearing on street, we assign a weight for every edge to reflect the density of buses on corresponding street.

● Each vertex $v_i \in \mathbb{V}$ refers to the intersection $i$ in a real road map.

● Each edge $e_{i,j} \in \mathbb{E}$ refers to the street $r_{i,j}$ between the intersection $i$ and $j$. Note that $e_{i,j} = e_{j,i}$.

● Each weight $\omega_{i,j} \in \mathbb{W}$ represents the weight of edge $e_{i,j} \in \mathbb{E}$, which reflects the density of buses on the corresponding street. The formula of weight is given as follows.

$$\omega_{i,j} = \begin{cases} \dfrac{1}{P_{i,j}}, & \text{if } P_{i,j} \neq 0 \\ \infty, & \text{otherwise} \end{cases} \quad (12)$$

The higher weight of the edge represents the density of buses on the corresponding street smaller. For convenience, the weight of the edge without bus lines passing by is set to a large constant, namely $\omega = \infty$. Since the movement of bus is bi-directional, we can assume that $\omega_{i,j}$ is equal to $\omega_{j,i}$.

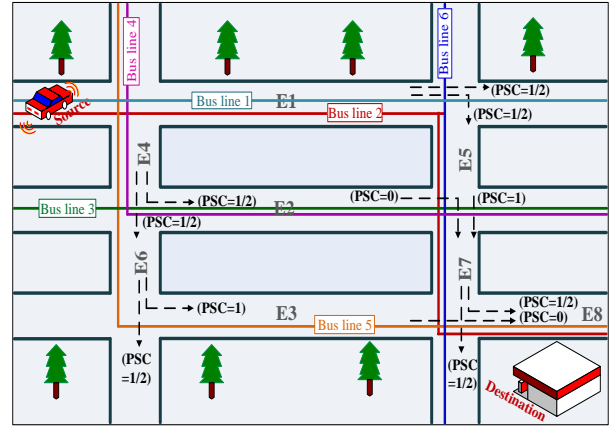

Fig.2. Analysis on the trajectory of bus lines

### B. Selection of Routing Path

In the routing with the mechanism of carry-and-forward, the higher density of vehicles along the routing path, the better the routing performance. Based on our proposed routing graph, we choose a sequence of streets between sender and destination with high density of buses as the routing path. In Fig. 2, there are three available path from the source to the destination: $P_1$ (*E1, E5, E7, E8*), $P_2$ (*E4, E2, E7, E8*) and $P_3$ (*E4, E6, E3, E8*). Among them, $P_1$ and $P_2$ have higher density of buses than $P_3$, thus are better as the routing path. Note that there is a challenge in the routing based on carry-and-forward mechanism: the transmission direction of the packet is affected by the driving direction of the carrier mostly. If the relay bus with packets fails to find the next available relay and to forward packet before it deviates from the routing path, the transmission direction of packets will also deviate from the routing path so that causes packet loss or re-routing. To solve the aforementioned problem, we propose a novel concept called the probability of street consistency (PSC), which is used to describe the consistency of bus lines between two adjacent streets. Based on PSC, we introduce the probability of path consistency (PPC) to evaluate routing paths. The path with higher PPC means smaller probability of buses deviating from this path, which is a better candidate as routing path.

*Definition* 3 (probability of street consistency, PSC): For any two adjoining streets $i$ and $j$, the amount of bus lines which pass through street $i$ is $N_i$, the amount of bus lines which pass through street $j$ is $N_j$, and the amount of bus lines which pass through street $i$ and street $j$ is $n_{i,j}$, the probability of consistency from street $i$ to street $j$ and the probability of consistency from street $j$ to street $i$ are respectively derived as follows:

$$\begin{cases} PSC_{i,j} = \dfrac{n_{i,j}}{N_i} \\ PSC_{j,i} = \dfrac{n_{j,i}}{N_j} \end{cases} \quad (13)$$

where $n_{i,j} = n_{j,i}$. Note that these formulas apply only to the condition that street $i$ adjoins with street $j$.

*Definition* 4 (probability of path consistency, PPC): A path consists of a sequence of streets, including $s_k, s_{k+1}, \cdots, s_n$ (for any $i$ ranging from $k$ to $n-1$, street $s_i$ adjoins with street $s_{i+1}$), the PPC of this path is derived as follows.

$$PPC(path) = \prod_{i=k}^{n-1} PSC_{i,i+1} \quad (14)$$

In consideration to the problem that the street whose PSC is zero will make the path's PPC equal zero, causing to ignore the contributions of streets which has non-zero PSC, we redefine the PPC, as follows.

$$PPC(path) = \frac{1}{n-k} \times \sum_{i=k}^{n-1} PSC_{i,i+1} + \prod_{i=k}^{n-1} PSC_{i,i+1} \quad (15)$$

In Fig. 2, $PSC_{1,5} = 1/2$, $PSC_{5,7} = 1$, $PSC_{7,8} = 1/2$, $PSC_{4,2} = 1/2$, $PSC_{2,7} = 0$, thus, $PPC(P_1) = 11/12$, $PPC(P_2) = 1/3$. Due to $P_1$ has higher value of PPC than the $P_2$, $P_1$ is supposed to be the routing path.

In this paper, there are two metrics to choosing routing path, including the density of buses along path and the probability of path consistency. Based on our proposed routing graph, we will choose the path with minimal sum of weights and maximal PPC as the routing path. Firstly, $k$ paths with minimal sum of weights will be found over the routing graph by using shortest path algorithm. Then, the PPCs of these $k$ paths are calculated by using Equation (15). The path with maximal PPC is selected as the routing path. The detail of the selection of routing path is shown as algorithm 1.

---

**Algorithm 1:** Selection of Routing Path

---

$S$ = the position of source bus;

$D$ = the position of destination;

$P$ = the set of available routing path.

***Bus:***

1: Compute shortest path from $S$ to $D$ by using shortest path algorithm;

2: Put all the shortest paths into $P$;

3: **for** each path in $P$ **do**

4:   Compute the PPC of path by using (15);

5: **end for**

6: Choose the path with maximum PPC as the routing path;

7: **end**

*C. Bus-Based Forwarding Strategy with ACO (FACO)*

In the process of routing, the relay bus that carries packets should select next relay bus along the routing path, ensuring that packets are delivered along their routing path to their destination. Only the bus which meets the qualification (shown as definition 5) can be the candidate of next relay bus.

*Definition* 5 (the qualification for the next relay, QR): If the routing path of packet is a sequence of $n$ streets (called $S$), and the carrier of packet is moving along the $i$-th street in $S$ (namely $S[i-1]$), the next relay must satisfy the condition that the street where it locates is between $S[i-1]$ and $S[n-1]$ in $S$.

Based on definition 5, the bus that carries packets can make the qualification for next relay according to the street where it locates and the routing path. If there are available and qualified buses in the communication range of carrier of packets, it will choose the bus with maximal link life time as the next relay. Otherwise, the carrier of packets will try to find eligible buses out of its communication range by multi-hop links. Considering that the multi-hop link could be broken down because of the movement of intermediate vehicle nodes, we introduce ant colony optimization algorithm (ACO) to seek a steady and long-lasting multi-hop link between carrier and the next relay bus. In FACO, buses work as main relay to route packets, and common cars work as secondary relay to build multi-hop link between buses.

*1) ACO problem formulation*

In this paper, ACO is used to seek qualified next relay bus and build a steady and reliable multi-hop link between two relay buses. The link life time and the transmission delay of multi-hop link are introduced as the metrics for evaluating multi-hop link. The higher link life time and the lower delay mean that the multi-hop link is more reliable. The multi-hop link between the carrier of packets (says $S$) and the candidate of next relay bus (says $B$) is made up of a set of common car nodes. We denote the multi-hop link between $S$ and $B$ as $L = \{S, c_1, c_2, ..., c_k, B\}$, where $c_1 \sim c_k$ stand for these secondary relays between $S$ and $B$. In $L$, there is a single-hop link between any two adjacent nodes. The establishment of the multi-hop link between two main relays can be modeled as an optimization problem with the following objective function.

$$\max F(L) = \varphi \times \frac{LT(L)}{1 + LT(L)} + (1-\varphi) \times \frac{1}{1 + D(L)} \quad (16)$$

where

$$\begin{cases} LT(L) = \min\left\{LT(l_{S,c_1}), LT(l_{c_k,B}), \min_{i \in [1,k-1]}[LT(l_{c_i,c_{i+1}})]\right\} \\ D(L) = D(l_{S,c_1}) + D(l_{c_k,B}) + \sum_{i \in [1,k-1]} D(l_{c_i,c_{i+1}}) \end{cases} \quad (17)$$

Subject to

$$D(L) \leq D_{th} \quad (18)$$

Where $\varphi \in (0,1)$ is a weight parameter which reflects the influence degree of the link life time on objective function, and $1 - \varphi$ refers to the influence degree of the delay on objective function. $LT(L)$ and $D(L)$ respectively denote the link life time and the transmission delay of multi-hop link $L$. Symbol $l_{c_i,c_j}$ denotes the single-hop link between node $c_i$ and node $c_j$. $LT(l_{c_i,c_j})$ and $D(l_{c_i,c_j})$ respectively denotes the link life time and the delay of the single-hop link between node $c_i$ and node $c_j$, where $LT(l_{i,j})$ can be deduced by using Equation (6). $D_{th}$ denotes the delay threshold.

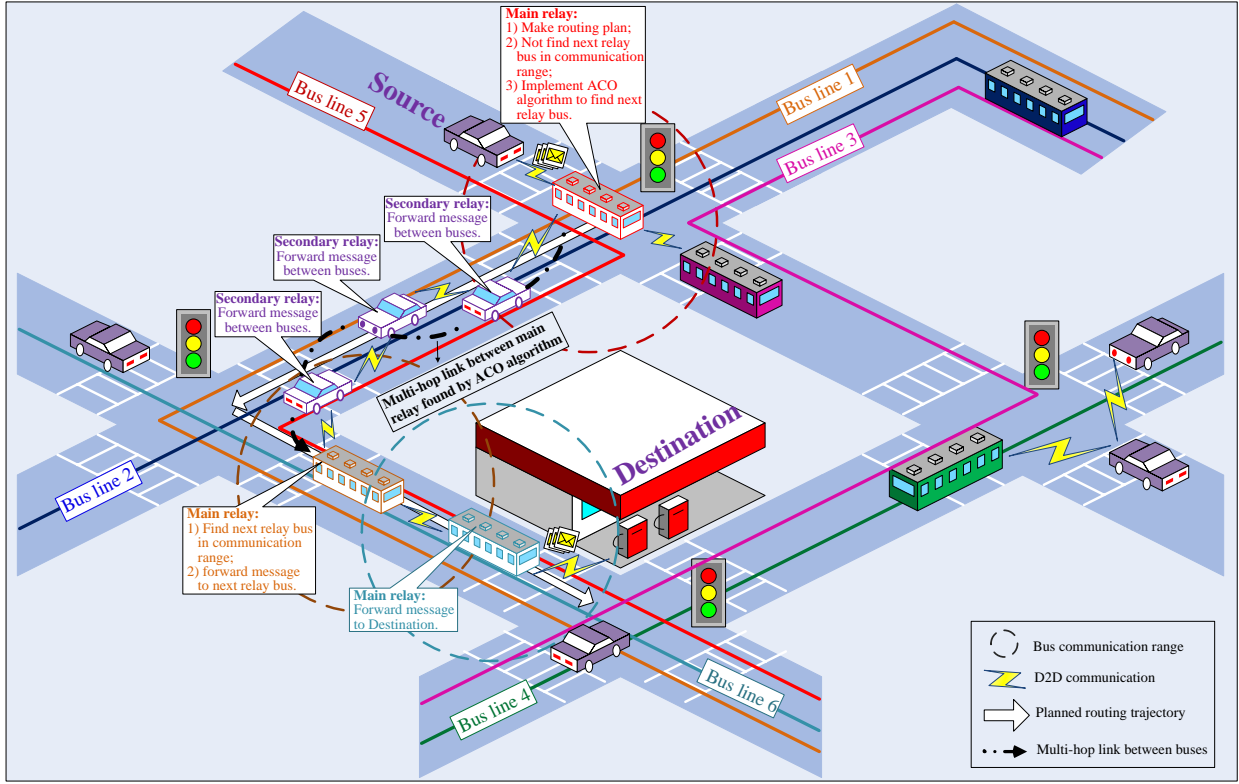

Fig.3. Routing scenario

*2) ACO establishment*

Every vehicle node stores pheromones and heuristic function values so that ant colony can finally find the optimal candidates of the next relay bus and the optimal multi-hop link between two relay buses. The establishment of ACO is divided into four phases, including request phase, discovery response phase and selection phase.

*Request phase*: The source bus $S$ which requests to find the next relay bus generates ask ants (AAs) and sets a timer for receiving response ants (RA), only receiving response ants that arrive within a valid period. Every AA carries the qualification for the next relay bus and a routing table used for recording the ID of nodes in multi-hop link. Here, the amount of AA is set as $N_{ant}$, and the timer is set as $D_{th}$. According to the pheromones and heuristic function values recorded in bus $S$, $N_{ant}$ AAs are respectively and stochastically sent to neighbor nodes by bus $S$ based on the probabilities given as Equation (19).

*Discovery phase*: Every vehicle $i$ which receives AA first records its ID into the routing table of AA. If vehicle $i$ is unqualified as the next relay bus, it will stochastically forward AA to its neighbors according to the probabilities (see Eq.(19)) during the life time of AA, or drop AA when the life time of AA expires. Otherwise, the eligible vehicle $i$ will generate the RA related to the AA. The probability shown below is used for vehicle $i$ to decide the forwarding direction of AA.

$$p_{i,j}(t) = \begin{cases} \dfrac{[\tau_{ij}(t)]^{\alpha} \times [\eta_{ij}(t)]^{\beta}}{\sum\limits_{k \in \mathbf{N}(i)} \left\{ [\tau_{ik}(t)]^{\alpha} \times [\eta_{ik}(t)]^{\beta} \right\}}, & if \ j \in \mathbf{N}(i) \\ 0, & otherwise \end{cases} \quad (19)$$

Where $\alpha$ and $\beta$ are respectively the factors of pheromone and heuristic function, they reflect the importance degrees of the residuary pheromone and heuristic function. $\mathbf{N}(i)$ is the set of neighbor nodes of the node $i$. $\tau_{ij}(t)$ denotes the pheromone intensity stored in node $i$ at time t. $\eta_{ij}(t)$ denotes the value of heuristic function at time $t$, which reflects the state of the link between node $i$ and its neighbor node $j$, as follows:

$$\eta_{ij} = \varphi \times \frac{LT(l_{i,j})}{1 + LT(l_{i,j})} + (1-\varphi) \times \frac{1}{1 + D(l_{i,j})} \quad (20)$$

*Response phase*: When the AA arrives at an eligible bus, a corresponding RA will be generated by this bus. RA carries a routing table recorded in AA, which represents the multi-hop link found by AA. RA will return to bus $S$ along the multi-hop link in reverse. Every vehicle which receives RA will forward RA to the next vehicle recorded in RA's routing table until RA arrives at the bus $S$. It will then update the value of heuristic function and the pheromone intensity by using equations (20) and (21) respectively.

$$\tau_{ij} = (1-\delta) \times \tau_{ij} + \delta \times \eta_{ij} \quad (21)$$

Where $\delta \in (0,1)$ is a weight parameter, which reflects the influence degree of heuristic function on the update of pheromone. Pheromones saved in every node will evaporate over time [23,49]. Considering the movement of vehicles, pheromones should evaporate to the initial state before the links expire to avoid the next AAs using the invalid links [49]. Thus, the pheromone will be close to the initial value if there is no RA passing by this link for a long time. Pheromones evaporate as follows:

$$\tau_{ij}(t+\Delta t) = \begin{cases} (1-\rho_{ij}(t)) \times \tau_{ij}(t), & if\ \tau_{ij}(t) > \tau_0 \\ \tau_0, & otherwise \end{cases} \quad (22)$$

Where $\tau_0$ is the initial value of pheromone, $\rho_{ij}(t) \in (0,1)$ denotes the pheromone evaporate coefficient at time $t$. $\tau_{ij}(t)$ and $\tau_{ij}(t+\Delta t)$ respectively denote the pheromone intensity at time $t$ and time $t + \Delta t$. Every other $\Delta t$, pheromones will evaporate as equation (22). When the life time of link $l_{i,j}$ expires, pheromones evaporate to their initial value, namely

$$\tau_0 = (1-\rho_{ij}(t))^\theta \times \tau_{ij}(t) \quad (23)$$

where $\theta$ denotes the evaporating number of pheromones within the link life time. Thus, $\theta$ and $\rho_{i,j}(t)$ are derived as follows:

$$\theta = \frac{LT(l_{i,j})}{\Delta t} \quad (24)$$

$$\rho_{ij}(t) = 1 - \sqrt[\theta]{\frac{\tau_0}{\tau_{ij}(t)}} \quad (25)$$

*Selection phase*: During the valid time period, bus *S* will save every received RA. Then it will use equations (16-18) to calculate the objective function values of the links recorded in these RAs. When the timer for receiving RAs beeps, bus *S* chooses the end node of the multi-hop link with maximal objective function value as the next relay bus.

In this paper, the bus-based forwarding strategy with ACO is summarized in Algorithm 2 and Algorithm 3.

---

*Algorithm* **2:** FACO Algorithm
---
QR = the qualification of next relay bus;
NR = the next relay bus;
***Candidates*** = the set of available next relay bus.
***Bus:***
1: **for** each packet arriving at $B_i$ **do**
2:   **for** each $b_i$ in $B_i$'s *NeighborTable* **do**
3:     **if** $b_i$ meets the condition of QR **then**
4:       Put $b_i$ into ***Candidates***;
5:     **end if**
6:   **end for**
7:   **if** *Candidate* is NULL **then**
8:     Find the next relay bus by using ***Ant Colony Optimization Establishment*** (see algorithm 3);
9:   **else**
10:     NR = the bus which has the maximum Link Lifetime with $B_i$ in ***Candidates***;
11:   **end if**
12:   Forward packet to NR;
13: **end for**

---

*Algorithm* **3:** Ant Colony Optimization Establishment
---
QR = the qualification of next relay bus;
TTL = $D_{th}/2$, the Time to Live of Ant;
***Path Sets*** = the set of available multi-hop path between buses.
***Source Bus:***
1: Generate $N_{ant}$ Ask Ants (AAs) with QR;
2: Send AAs to its neighbors based on the probability in (19);
3: Wait();
***Vehicle $V_i$:***
4: **for** each AA arriving at $V_i$ **do**
5:   Put $V_i$'s id into the relay table of AA;
6:   **if** the TTL of AA expires **then**
7:     Drop AA;
8:   **else**
9:     **if** the type of $V_i$ is bus and $V_i$ meets the condition of QR **do**
10:       Generate corresponding Response Ant(RA) with the relay table of AA;
11:       Send RA to the next vehicle recorded in the relay table of RA reversely;
12:     **else**
13:       Forward AA to find available relay bus;
14:     **end if**
15:   **end if**
16: **end for**
17: **for** each RA arriving at $V_i$ **do**
18:   Update heuristic function $\eta_{ij}$ by using (20);
19:   Update pheromone $\tau_{ij}$ by using (21);
20:   Send RA to next node recorded in the relay table of RA;
21: **end for**
***Source Bus:***
22: **while** the duration of wait() is less than $D_{th}$ **do**
23:   **for** each RA arrives **do**
24:     Update heuristic function $\eta_{ij}$ by using (20);
25:     Update pheromone $\tau_{ij}$ by using (21);
26:     Compute the value of objective function by using (16-17);
27:     Put the multi-hop link recorded in the routing table of RA and the value of the corresponding objective function into ***PathSets***;
28:   **end for**
29: **end while**
30: Choose the end node of the multi-hop link with maximum value of the objective function as the next relay bus;
31: Send packet to next bus along the selected multi-hop link.
32: **end**

## V. PERFORMANCE EVALUATION AND ANALYSIS

In this section, we describe the simulation environment and show the simulation results.

### A. Simulation Environment

We implement our simulations by using the network simulator (OMNeT++ 4.4.12) [50] and vehicular network simulation framework (Veins 4.6) [51] on win7 (64 bits). The road traffic simulator (SUMO 0.30.0) [50] is used to generate the routes of buses and the moving trajectories of common cars. The simulation area is set as the map inside the first ring road of Chengdu which is downloaded from OpenStreetMap [53]. The simulation map consists of 58 intersections and 100 streets (see Fig. 4 (a)). In our simulations, there are altogether 400 buses running on 20 bus lines (see Fig. 4 (b)). According to the number of common cars (namely non-bus) in the network, we classify the simulation scenarios into three categories: sparse network, common network and dense network which respectively contain 2000, 4000 and 6000 common cars. Buses drive along their own fixed bus line, and common cars randomly distribute and run on roads. The maximum driving velocity of vehicles is 40 kilometers per hour. And other simulation parameters are given in Table I.

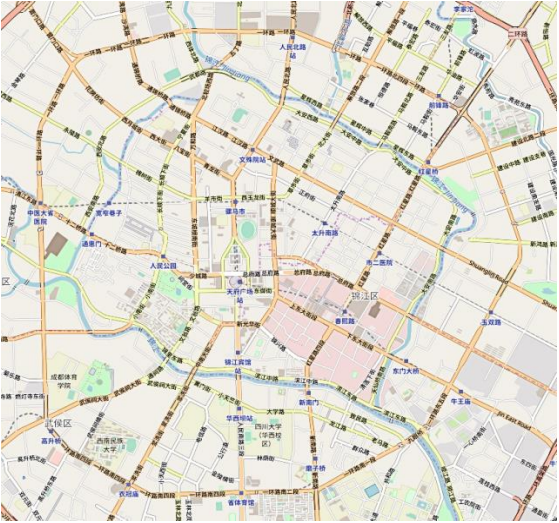

(a) The real map inside the First Ring Road of Chengdu in China

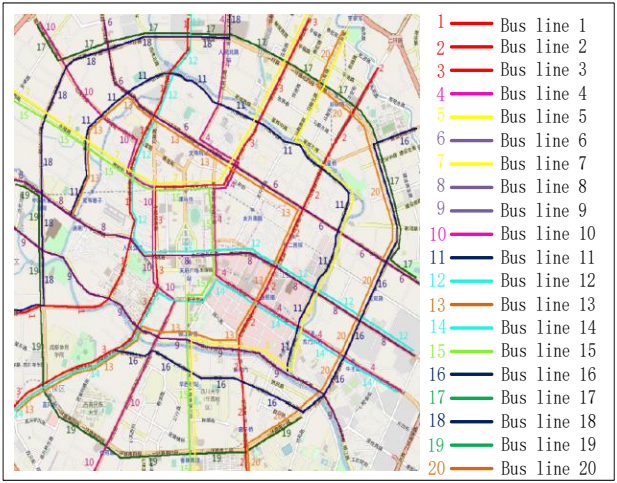

(b) The trajectories of 20 bus routes

Fig.4. The map and the bus lines used in simulation

To study the performance of our proposed routing algorithm, we introduce two existing routing algorithms for comparisons, i.e., CBS [23] and AQRV [54]. Based on the bus network, CBS adopts a node-centric routing scheme, and its main idea is selecting a group of buses which encounter with each other frequently as routing path. It builds the bus-based bone graph by analyzing the trajectories of bus lines, and all bus lines are divided into 6 communities by implementing the community detection technology. In the bus-based bone graph, every vertex denotes a bus line and every edge reflects the bus lines corresponding with the two vertexes of this edge. They will then contact each other. The weight of every edge represents the probability of contact between two bus lines corresponding to these two vertexes. An optimal routing path is selected by using shortest path algorithm over the bus-based bone graph. AQRV aims at achieving the optimal QoS of routing. It utilizes the ant colony optimization algorithm to find an optimal routing path with best delivery ratio, higher connectivity probability, and lower delivery delay. It can dynamically select a routing path according to real time network condition.

TABLE I
SIMULATION PARAMETERS

| Parameter | Setting |
|---|---|
| Simulation map range | About $5.9 \times 5.8 km$ |
| MAC protocol | IEEE 802.11p |
| Communication radius | $200 \sim 800 m$ |
| Transmission rate | $6 Mbps$ |
| Vehicle velocity | $10 \sim 40 km/hr$ |
| Number of bus lines | 20 |
| Ask Ant number $N_{ant}$ | 10 |
| Delay upper threshold $D_{th}$ | $10s$ |
| FACO parameters | $\tau_0 = 0.3, \delta = 0.7, \varphi = 0.6,$ $\alpha = 8, \beta = 5, \Delta t = 1s$ |
| Simulation duration | $4000s$ |

In this paper, two metrics, including the packet transmission ratio and the average end-to-end delay are used to evaluate the routing performance.

1) *packet transmission ratio*: It is defined as the ratio of the amount of packets transmitted successfully to the total amount of packets generated, which reflects the reliability of routing algorithms [55].

2) *average end-to-end delay*: It is defined as the ratio of the sum of transmission time of all the packets transmitted successfully to the amount of packets transmitted successfully, which reflects the efficiency of routing algorithms [56].

### B. Simulation Results

To analyze the routing algorithm's adaptability to networks with different vehicle densities, we conduct simulations in three network scenarios, i.e. sparse network, common network and dense network, where the number of vehicles are 2000, 4000 and 6000 respectively. In every scenario, we test the changes of the packet transmission ratio and the average end-to-end delay under the effects of different transmission distance and communication radius. Every scenario is divided into four groups in which the communication radiuses are respectively set as 200 meters, 400 meters, 600 meters and 800 meters. In every group, the packet transmission distance ranges from 0 to 2500 meters.

In the sparse network, there are 400 buses and 2000 common cars running on our simulations. With the changes of two parameters, including the packet transmission distance and the communication radius, the changes of the packet transmission ratio and the average end-to-end delay are shown as Fig 5 and Fig 6 respectively.

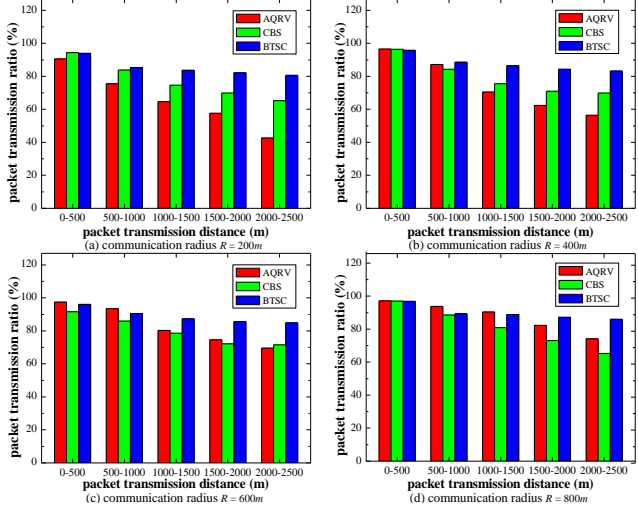

Fig.5 The packet transmission ratio in sparse network which has 200 buses and 2000 common cars

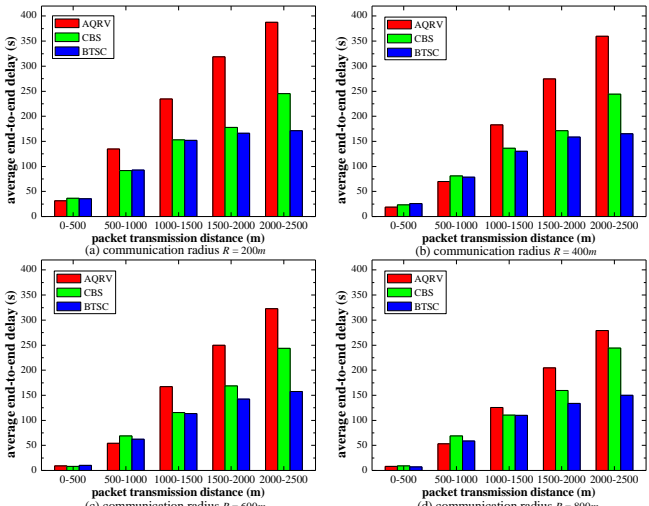

Fig.6 The average end-to-end delay in sparse network which has 200 buses and 2000 common cars

We can see that every algorithm has similar change trends in these four groups. With the transmission distance increase, the packet transmission ratios of these three algorithms are on a declining trend and the average end-to-end delays of these three algorithms are on uptrend in these four groups. Among these three algorithms, BTSC rises less sharply in the average end-to-end delay and falls more slightly in the packet transmission ratio than other two. This reflects our proposed routing algorithm has a stable performance on different transmission distance. With the communication radius increase, all these three algorithms have improvements on the routing performances which reflect in the increasing packet transmission ratio and the decreasing average end-to-end delay. In the case of same packet transmission distance, the bigger the communication radius, the less the amount of the forwarding hop during packet routing. As the amount of forwarding hop decreases, the success rate for routing packet grows and the transmission delay gets lower. Therefore, the increasing communication radius makes the improvement in routing performance of these three algorithms.

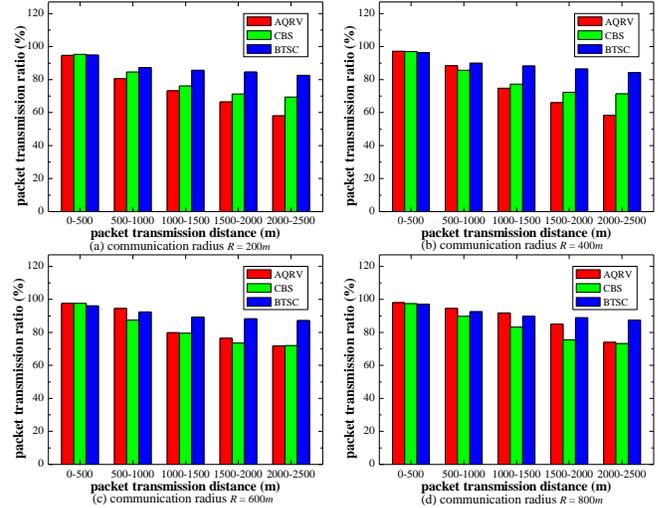

Fig.7 The packet transmission ratio in common network which has 200 buses and 4000 common cars

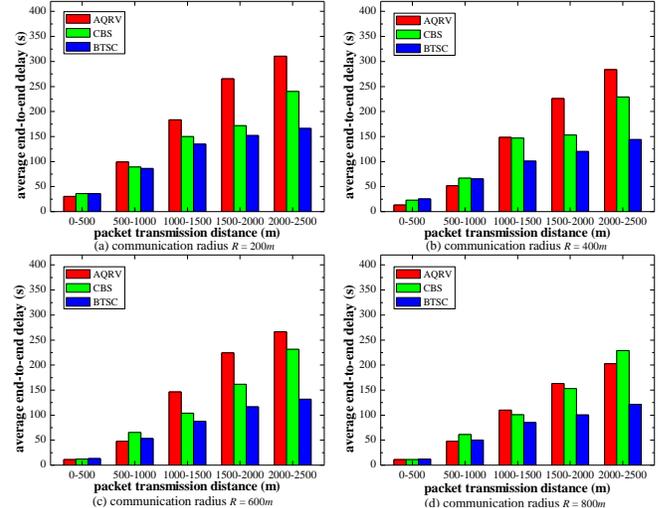

Fig.8 The average end-to-end delay in common network which has 200 buses and 4000 common cars

In the common network, the number of buses is 400, and the number of common cars increases to 4000. Fig 8 and Fig 9 present the changes of packet transmission ratio and the average end-to-end delay under the effects of different packet transmission distance and communication radius. In Fig 8, the packet transmission ratios of these three algorithms decrease with packet transmission distance increases but improve with communication radius increase. The packet transmission ratio of BTSC presents slowly descending tendency when the packet transmission distance exceeds 1500 meters. This means that our proposed algorithm has good convergence for packet transmission distance. On the contrary, the average end-to-end delays of these three algorithms get higher with the packet transmission distance increases in Fig 9. From Fig 8 and Fig 9, we can see that these three algorithms have similar routing performances that the packet transmission ratios of all the three are over 90 percent and the average end-to-end delays are below 25 seconds when the packet transmission distance is between 0 and 500 meters. It is because that the packets mostly are delivered in the form of single-hop link when the packet transmission distance is no more than the communication radius, resulting in greater routing performance. When the packet transmission distance exceeds 1000 meters, our proposed algorithm has the best performance among these three algorithms.

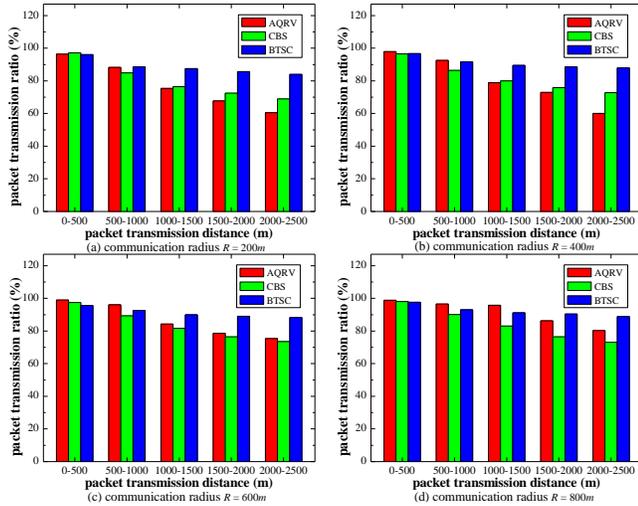

Fig.9 The packet transmission ratio in dense network which has 200 buses and 6000 common cars

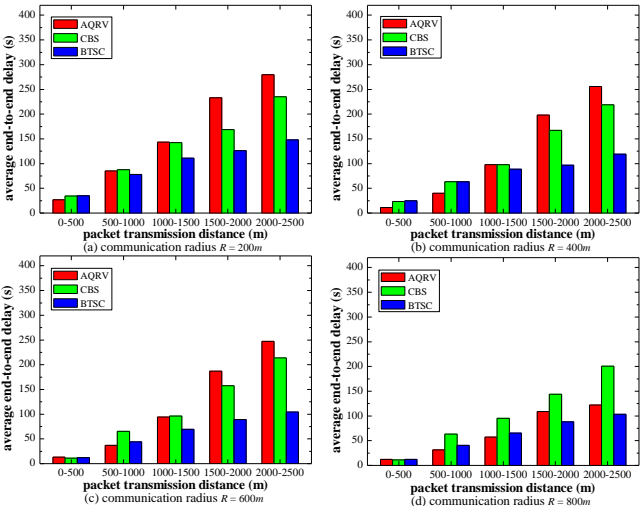

Fig.10 The average end-to-end delay in dense network which has 200 buses and 6000 common cars

In the dense network, there are fifteen times more common cars than buses, and the number of common cars is up to 6000. With the packet transmission distance and communication radius change, the changes of the packet transmission ratios and the average end-to-end delay are shown in Fig 9 and Fig 10. Like the results of sparse network and common network, the routing performance improves with the increasing of communication radius or the decreasing of packet transmission distance. Unlike the results of preceding two networks, AQRV has better performance than other two algorithms when the packet transmission distance is no more than 1000 meters. Since the ant colony optimization utilized in AQRV can quickly find next relay in the condition of high vehicle density. Although BTSC adopts the scheme of ant colony optimization, the scheme of two kind relays used in our proposed algorithm where bus is main relay and common car is secondary relay makes the performance of BTSC inferior to AQRV's when the packet transmission distance is short. However, BTSC has good convergence and stability with the packet transmission distance changes because of two kind relays used in BTSC.

By comparing the routing performances in these three networks, there are some common features of the effect of packet transmission distance and communication radius on routing performances in these three networks, shown as follows.

1) With the transmission distance increase, the packet transmission ratios of these three algorithms are on a declining trend and the average end-to-end delays of three algorithms are on uptrend in these three networks.

2) With the communication radius increase, the packet transmission ratio and average end-to-end delay of these three algorithms have improvement in these three networks.

3) With the transmission distance change, BTSC has good convergence and stability.

Above all, we can figure out that the effects of packet transmission distance and communication radius on performances are similar in these three networks, and our proposed routing algorithm has a more stable performance under the effect of packet transmission distance and communication radius in these three networks.

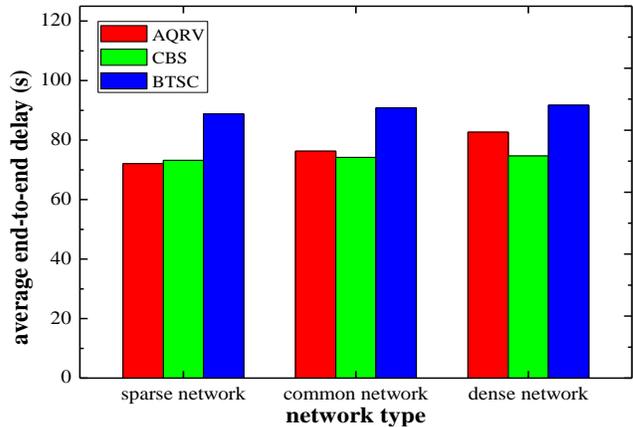

Fig.11 The changes of packet transmission ratio with different vehicle densities

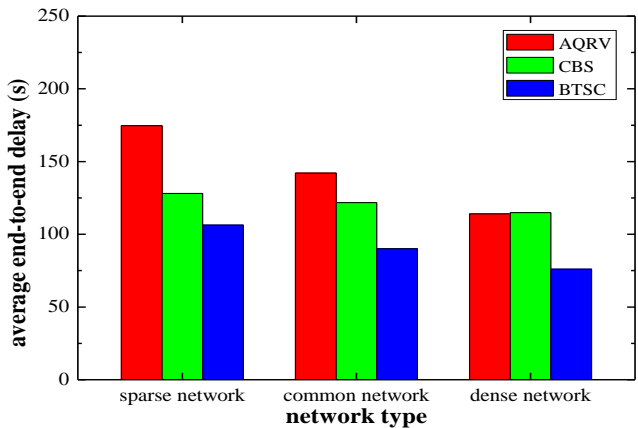

Fig.12 The changes of average end-to-end delay with different vehicle densities

To compare the adaptabilities of these three routing algorithms to networks of the different node densities, we analyze the effects of car density on the transmission ratio and the average end-to-end delay in these three algorithms. The communication radiuses are randomly selected from 200 to 800 meters, and the packet transmission distances are randomly selected from 0 to 2500 meters in these three networks where the number of buses are 400 and the numbers of common cars are 2000, 4000 and 6000. Fig 11 and Fig 12 present the changes of packet transmission ratio and average end-to-end delay with the density of nodes. As can be seen from Fig 11, the packet transmission ratios of AQRV and our proposed algorithm are on the rise as the number of common cars increases and CBS's is relatively stable. In Fig 12, the average end-to-end delay decreases with the density of vehicle nodes increasing in AQRV and BTSC but remains unchanged in CBS. It's because that both AQRV and our proposed algorithm utilize common cars as the relay to forward packets. Therefore, the increase of the density of common cars is beneficial to routing performance. The AQRV's routing performances in terms of average end-to-end delay and packet transmission ratio have more dramatic change than our proposed algorithm because it is based on the superiority of buses. Buses have fixed driving trajectories and the number of buses is a nearly constant. As a result, the routing performance of bus-based routing algorithms is not susceptible to the impact of the number of common cars. In these three networks, our proposed algorithm has better performances in the average end-to-end delay and the packet transmission ratio than other two algorithms.

By comparing the performances of these three algorithms under the effects of packet transmission distance, communication radius and the density of vehicle nodes, we can see that our proposed routing algorithm has a better adaptability to networks with different vehicle density, better performances in terms of the packet transmission ratio and the average end-to-end delay, and good convergence for the packet transmission distance.

## VI. Conclusion

In this paper, we have proposed a novel bus-based street-centric routing algorithm (BTSC), which adopts the street-centric scheme. Since buses have fixed routes, we build a bus line-based routing graph by analyzing the probability of buses appearing on every street. We propose two novel concepts, i.e. PSC and PPC, which are used as the metrics to select routing path, aiming at choosing a routing with the higher density of buses and the smaller probability of transmission direction deviating from routing path. Buses on the selected routing path play the role of main relay to delivery packets to destination. Considering relay buses have difficulty finding the next available hop relay within its communication range in practice, we propose a bus forwarding strategy with ACO (FACO), which employs ACO algorithm to find a reliable and steady multi-hop link between two relay buses which cannot directly communicate with each other. This aims to increase the chances of bus forwarding and decrease end-to-end delay. In this paper, the street-centric routing algorithm is optimized from two aspects, including the selection of routing path and the forwarding strategy, to improve QoS in terms of transmission ratio and delay. The simulation results show that our proposed routing algorithm has a better performance in terms of transmission ratio and average end-to-end delay and has a better adaptability to networks of different vehicle density.